\documentclass[prl,twocolumn,showpacs,floatfix,amsmath,amssymb,superscriptaddress] {revtex4}
\usepackage[dvips]{graphicx}
\usepackage{latexsym}
\usepackage{graphicx}
\usepackage{times}
\usepackage{amsmath}
\usepackage{dcolumn}
\usepackage{latexsym,amsmath,amssymb,bm,euscript}
\def\CsCuCl{$\text{Cs}_2\text{CuCl}_4$}
\def\CsCuBr{$\text{Cs}_2\text{CuBr}_4$}

\def\DM{\text{Dzyaloshinskii-Moriya}}
\def\beq{\begin{equation}}
\def\eeq{\end{equation}}
\def\bea{\begin{eqnarray}}
\def\eea{\end{eqnarray}}

\begin{document}

\title{Quantum stabilization of 1/3-magnetization plateau in Cs$_2$CuBr$_4$}

\author{Jason Alicea}
\affiliation{Department of Physics, California Institute of  
Technology, Pasadena, CA 91125}
\author{Andrey V. Chubukov}
\affiliation{Department of Physics, University of Wisconsin, Madison,  
WI 53706}
\author{Oleg A. Starykh}
\affiliation{Department of Physics, University of Utah, Salt Lake  
City, UT 84112}

\date{\today}

\begin{abstract}

We consider the phase diagram of a spatially anisotropic 2D triangular antiferromagnet in a magnetic field. 
Classically, the ground state is umbrella-like 
for all fields, but we show that the quantum phase diagram is
much richer and contains a 1/3 magnetization plateau, two commensurate planar states, two 
incommensurate chiral umbrella phases, and, possibly, a planar state separating the two chiral phases.
Our analysis sheds light on several recent experimental findings for the spin-1/2 system \CsCuBr. 

\end{abstract}

\maketitle

\begin{figure}
   \includegraphics[width=2.3in]{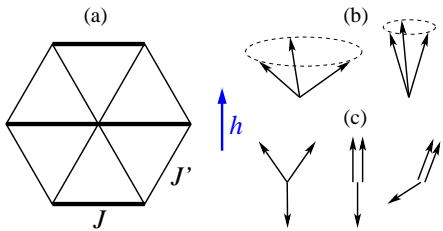}
   \caption{
     \label{GroundStates}(a) Anisotropic triangular lattice with horizontal exchange $J$ and diagonal exchange $J'$.  
     (b) Umbrella and (c) planar phases comprise competing classical ground states of the isotropic nearest-neighbor model.}
\end{figure}

\emph{Introduction.} ~A defining characteristic of frustrated quantum magnets  
is the appearance of numerous competing orders.  This competition dramatically enhances quantum fluctuations, generating highly non-classical behavior as exemplified by, \emph{e.g.}, \CsCuCl~and \CsCuBr. 
These materials comprise quasi-2D spin-1/2 triangular antiferromagnets
with spatially anisotropic exchange [see Fig.\ \ref{GroundStates}(a)] and weak \DM~(DM) coupling.
Absent the latter, both systems classically should realize a zero-field 
coplanar spiral, which evolves into non-coplanar ``umbrella'' states in 
a field as in Fig.\ \ref{GroundStates}(b) with smoothly increasing magnetization up to saturation \cite{veillette05}.  Experiments, however, reveal 
 decidedly different, non-classical  behavior:   
in fields directed along the triangular layers
\CsCuCl~realizes \emph{commensurate} 
coplanar order in a wide field range
with smoothly increasing magnetization
 ~\cite{tokiwa06,veillette06}, and \CsCuBr~
 exhibits collinear ``up-up-down'' (UUD) order shown in Fig.\  \ref{GroundStates}(c) over a finite field interval,
yielding a 1/3 magnetization plateau~\cite{ono03,ono04,ono05,takano07}. 
 Neither observation is accounted for within a classical analysis \cite{veillette05}.
NMR \cite{takigawa04,takigawa07} and neutron scattering \cite{ono05} 
additionally find planar states
adjacent to the UUD phase, with neutron and thermodynamic measurements \cite{takano07} 
indicating that the transitions are first order.  
Additional experiments \cite{ono05,takano08} on \CsCuBr~also suggest  
the presence of a narrow 2/3-plateau and additional intervening collinear phases near particular fields.

\begin{figure}
   \includegraphics[width=2.5in]{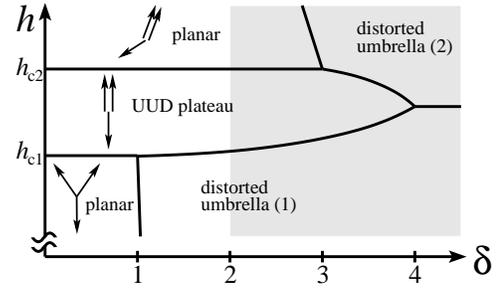}
   \caption{
     \label{PhaseDiagram}Proposed phase diagram for the anisotropic nearest-neighbor Heisenberg model near 1/3 magnetization (full field range not shown).  
     The horizontal axis is $\delta = (40/3)S(J-J')^2/J^2$.  Planar states shown are \emph{commensurate}, though they are expected to be incommensurate at small and large fields.
The shaded area is where the UUD and adjacent phases are metastable, the energy being minimized by umbrella states of Fig.\ \ref{GroundStates}(b).}
\end{figure}

While the existence of the UUD phase is well-established for the isotropic triangular antiferromagnet, much less is known about the stability of the plateau and the proximate quantum phases in the anisotropic case.  
The challenge here is illuminated by first observing that in the isotropic limit, the UUD state appears due to an ``accidental'' classical degeneracy between umbrella and planar states shown in Figs.\ \ref{GroundStates}(b) and (c), 
which quantum fluctuations lift in favor of the latter \cite{chubukov91}. 
When  $J' \neq J$, however, 
this degeneracy is lifted already at the classical level,
but in favor of \emph{umbrella} states for all fields. The planar 
phases can then only emerge if quantum effects overshadow those of spatial anisotropy. 
This in turn implies that the standard spin-wave expansion is not applicable 
since planar phases cease to be classical ground states.
To address the quantum phase diagram for the anisotropic system, particularly the phases near 1/3 magnetization, 
we introduce a modified approach here which is controlled by the smallness of  $1/S$ \emph{and} 
spatial anisotropy, and yields results which are non-analytic in both parameters.  

Figure \ref{PhaseDiagram} summarizes our results.
We find that the physics is controlled by the parameter $\delta = (40/3) S (J-J')^2/J^2$.  For $\delta <1$
 the stability of the UUD phase is, counter-intuitively, unaffected by anisotropy.  
 Moreover, the spin order remains coplanar and \emph{commensurate}
at fields both below and above the UUD phase; incommensurate phases appear only at small and high fields.
  For $1<\delta<4$, the UUD phase persists, but at the boundaries it becomes unstable 
  towards non-coplanar, incommensurate phases  
which can be regarded as 
 distorted umbrellas (this happens for $\delta >1$ at the lower boundary and
 for $\delta >3$ at the upper boundary). 
 These two phases emerge as finite-$k$ instabilities 
 of the two low-energy spin-wave branches of the UUD phase,
 and both have a non-zero Ising order parameter associated with
 chirality 
$K_{ABC} = \hat{z} \cdot ({\bf S}_A \times {\bf S}_B + {\bf S}_B  \times {\bf S}_C +  {\bf S}_C  \times {\bf S}_A)$
 for each plaquette. 
 For $\delta >4$, the UUD state ceases to exist, and there is no magnetization plateau.  Since the chiralities of the low- and high-field distorted umbrella phases are uncorrelated, the two must be separated by (at least) a first order transition in this region.
 At still larger $\delta$ (stronger anisotropy), 1D physics becomes important, and the
 system cannot be described by our semi-classical theory.

As a further complication, for $\delta >2$ the energy of the UUD state becomes larger than that of the classical, 
undistorted umbrella, \emph{i.e.}, for $2< \delta <4$, the UUD state and neighboring distorted umbrellas are 
{\em metastable}.
We represent this by shading the region $\delta >2$ in Fig.\ \ref{PhaseDiagram}.  We expect that these metastable phases may be 
probed in pulsed field experiments \cite{narumi04}.

\emph{Model and UUD state in the Anisotropic System.}~ 
We consider a simple Heisenberg model with 
\begin{equation}
   H = \sum_{\langle {\bf r r'}\rangle} J_{\bf r r'} {\bf S}_{\bf r} 
\cdot {\bf S}_{\bf r'} - h S \sum_{\bf r} S^z_{\bf r},
  \label{spinH}
\end{equation}
where ${\bf S}_{\bf r}$ are spin-$S$ operators, the exchanges $J_{\bf r r'}$ are as shown in Fig.\ \ref{GroundStates}(a), 
and $h$ is the  (scaled)
magnetic field. The saturation field  is $h_{\rm sat} = (2J + J')^2/J$.  Since we wish to treat quantum effects and the effects 
of anisotropy on equal footing, we will organize our analysis by assuming 
that both $(J-J')/J$ and $1/S$ are small. 

With $J = J'$, the two competing classically degenerate states are commensurate (three-sublattice) umbrella and 
planar states shown in Fig.\ \ref{GroundStates}. Quantum fluctuations favor 
planarity, and  spin re-arrangement in a field occurs as in Fig.\ \ref{GroundStates}(c). This process includes an intermediate UUD phase, which is
 classically 
stable only at $h_{sat}/3$, but quantum fluctuations extend its stability 
 to a finite field interval,
$h_{c1}^0 \leq h \leq h_{c2}^0$ \cite{chubukov91},
 resulting in a 1/3-magnetization plateau.
This is not surprising given that quantum fluctuations generally favor collinear states \cite{shender82,henley88}. 
To leading order in $1/S$ 
\begin{equation}
h_{c1}^0 = 3J - \frac{0.50 J}{2S} ~,~  h_{c2}^0 = 3J + \frac{1.3 J}{2S},
\label{eq:plateau1}
\end{equation}
which for $S = 1/2$ yields a plateau in a range
$\Delta h^0 = h^0_{c2}-h^0_{c1} = 1.8J/(2S)$, in good agreement with exact diagonalization \cite{honecker04}. Inside this range, there are two
 low-energy spin-wave modes with gaps $\propto |h^0_{c1,2} -h|$ at ${\bf k}=0$. 

 When $J \neq J'$, the
umbrella state becomes incommensurate, and classically has lower 
energy than the planar phase for all fields. The naive 
expectation, then, is that the UUD phase must immediately shrink and disappear as $|J-J'|$ increases. 
We show, however, that the actual situation is much more complex, with new phases emerging when $J \neq J'$. 

To study the stability of the classically unfavorable UUD state, we
explore a modified large-$S$ approach to Eq.\
 (\ref{spinH}). First, we introduce a three-sublattice
representation where spins on the A and B sublattices point up while
those on the C sublattice point down, and use the standard Holstein-Primakoff mapping. The usual linear 
spin-wave Hamiltonian obtained in this fashion is not an appropriate starting point 
due to the classical instability of harmonic spin waves at $\delta \neq 0$.  However, the \emph{interacting} spin-wave 
Hamiltonian must support a stable UUD plateau over a finite anisotropy range, as exact diagonalization finds \cite{miyahara06}.  Therefore, we extend the 
linear spin-wave Hamiltonian of the UUD state to include the leading $1/S$ self-energy corrections
obtained by decoupling the quartic interactions using correlations from the {\em isotropic} system.

Diagonalizing this Hamiltonian, we obtain three spin-wave branches. One branch describes a precession of the 
total magnetization and has a high energy $\sim h_0 \equiv J + 2J'$.
The energies of the other two branches are small near
${\bf k} = 0$: 
\begin{eqnarray}
  H_{\text{uud}} &=& S\sum_{\bf k}[ \omega_1 d^\dagger_{1,\bf k}d_{1,\bf k}  
+ \omega_2 d^\dagger_{2,\bf k} d_{2,\bf k}],
\label{eq:H-uud-prime}
\end{eqnarray}
where the leading expressions at small ${\bf k}$ are
\begin{eqnarray}
 \omega_{1,2}({\bf k}) &=&
 \pm \left(h-h_0-\frac{1}{5S}J-\frac{3}{4}J{\bf k}^2\right) + \frac{3J Z}{20S}
\label{eq:uud5}
\end{eqnarray}
with $Z =\sqrt{9 + 10 S [6 k_y^2 + 10 S k_x^4 - 3 k_x^2 (\delta-2)]}$.
The critical fields obtained from these energies are
\beq
 h_{c1,2} =  h^0_{c1,2} +2(J'-J) \mp \frac{3J}{4} {\rm min} \left( \mp k_x^2  +  \frac{Z-3}{5S}\right),
\label{hc12}
\eeq
where  $h^0_{c1/2}$ is given by (\ref{eq:plateau1}), and  
the minimum is taken with respect to $k_x$ ($k_y =0$ at the minimum  for all $\delta$). The UUD phase 
is stable for $h_{c1} < h < h_{c2}$.

These results, which are non-analytic in $1/S$ and $J-J'$, encode the physics governing the local stability of the  UUD state 
in the anisotropic system.   One can verify by sending $S \rightarrow \infty$ above that the UUD state is indeed 
unstable for any non-zero anisotropy in the classical limit, due to an instability at finite $k_x$.  
Surprisingly, in the quantum system a finite amount of 
anisotropy is required to begin destabilizing the plateau.  Specifically, for $\delta < 1$ both modes are 
minimized at ${\bf k} = 0$, so it follows from Eq.\ (\ref{hc12}) that the plateau width $\Delta h$ is unchanged 
from the isotropic system. The effect of anisotropy in this regime is only to shift the 
plateau's location and soften the dispersion around ${\bf k} = 0$.

For $\delta > 1$, the minimum of $\omega_1$ shifts to ${\bf k}_{1\pm} = (\pm k_1,0)$, where
$k_1^2 = [3\delta-6+\sqrt{3\delta(4-\delta)}]/(20S)$; the lower critical field then moves upward, 
reducing the width of the UUD plateau (see Fig.\ \ref{PhaseDiagram}). Similarly, for $\delta > 3$ the minimum of $\omega_2$ 
shifts to ${\bf k}_{2\pm} = (\pm k_2,0)$, with $k_2^2 = [3\delta-6-\sqrt{3\delta(4-\delta)}]/(20S)$. 
 At this point the upper critical field moves to a smaller value, 
further reducing the UUD region.
The plateau ceases to be locally stable at $\delta = 4$, when both
spin-waves become gapless at 
$k^2_1 = k^2_2 = k^2_{m} = 3/(10S)$. 

Let us now explore the phases that emerge immediately away from the UUD state.
At $h_{c1}$ and $h_{c2}$, magnons Bose condense, and one must 
determine the energetically favorable combination of operators $d_{1,2{\bf k}}$ that condenses, and
what this implies for the spin components $\langle S^{x,y}\rangle$.  
For $\delta <1$ this is straightforward:  the minima of $\omega_{1,2}({\bf k})$ occur
 at ${\bf k} = 0$ , and the
 order parameters are simply $\psi_{1,2} \propto \langle d_{1,2,{\bf 0}}\rangle$.
One can easily verify that condensation of $\psi_{1} (\psi_2)$  at $h=h_{c1} ~(h_{c2})$
leads to the \emph{commensurate} coplanar spin 
configurations
 displayed in Fig.\ \ref{GroundStates}(c).
The prediction of commensurate order adjacent to the UUD state 
over a range of anisotropy is rather nontrivial, and could 
 be tested in exact diagonalization studies.

The situation is subtler at the lower critical field when $\delta >1$, 
 since here $\omega_1({\bf k})$ possesses
two inequivalent minima at ${\bf k}_{1\pm}$.  There are then two order 
parameters, $\psi_\pm = \sqrt{3/NS} \langle d_{1,{\bf k}_{1\pm}} \rangle$ ($N$ is the number of spins), whose
energy derived from the interacting spin wave Hamiltonian \cite{nikuni95} is
\begin{equation}
\frac{2E}{JNS^2} = r(|\psi_+|^2 + |\psi_-|^2) + 
(|\psi_+|^2 + |\psi_-|^2)^2 + u |\psi_+|^2|\psi_-|^2.
\label{n_1}
\end{equation}
Here  $r\propto h-h_{c1}$ and $u = 2 \cosh^2 {2\phi_{k_1}}$, 
 where 
\beq
\tanh (2 \phi_{k_1}) = \frac{6 (J-J') k_1}{\omega_1 (k_1) + \omega_2 (k_1)} = \sqrt{3\delta} \frac{\sqrt{10S} k_1}{3 + 10 S k_1^2}.
\label{n_2}
\eeq
 Since $u>0$, 
below the transition interactions favor $\psi_+ \neq 0$, $\psi_- = 0$ or vice versa. Choosing the former, the spin configuration can 
be written $\langle S^+_{A}\rangle =
 - S \psi_+ (\cosh \phi_{k_1} + i \sinh \phi_{k_1}) e^{-ik_1 x}$, 
$\langle S^+_{B} \rangle = S \psi_+ (\cosh \phi_{k_1} + i \sinh \phi_{k_1}) e^{-ik_1 x}$, $\langle S^+_{C}
 \rangle = 2i S \psi_+ \sinh  \phi_{k_1} e^{+ik_1 x}$.
This corresponds to \emph{non-coplanar}, \emph{incommensurate} order that can be described 
as a distorted umbrella.  Non-coplanarity of this state leads to a finite chirality $K^{(1)}$,
the sign of which is determined by that of the condensate momentum via
$K^{(1)}_{ABC} = \pm 3 S^2 |\psi_{\pm}|^2  \sinh {2 \phi_{k_1}}$. 

The same consideration holds at the 
upper critical field when $\delta >3$: $\omega_2({\bf k})$ 
again has two inequivalent minima at ${\bf k}_{2\pm}$, and the energy has the same form as in (\ref{n_1}), 
with the order parameter ${\bar \psi}_\pm = \sqrt{3/NS} \langle d_{2,{\bf k}_{2\pm}} \rangle$.  The spin configuration above $h_{c2}$ is another
 distorted umbrella with chirality $K^{(2)}_{ABC} = \mp 3S^2 |{\bar \psi}_{\pm}|^2 
\sinh {2 \phi_{k_2}}$. 

At $\delta =4$, the UUD plateau shrinks to a point at $h_c = h_0 + 17J/(40S)$, 
and  becomes unstable at larger $\delta$. 
 How the two distorted umbrellas merge in this regime presents an interesting issue.  Since these states arise upon condensation of \emph{different} spin-wave modes at $h_{c1,2}$, their chiralities are uncorrelated.
The  two phases then 
cannot gradually transform into each other and must be separated either by a first order transition, 
or by an intermediate phase with no chirality. 

To gain insight here we study the instability of the UUD phase at $\delta =4, h = h_c$. At this point,
 the two spin-wave branches become gapless at the same 
 $\pm k_m$,  and the coherence factors $\sinh \phi_{k_{m}}$ and $\cosh \phi_{k_{m}}$ diverge as $1/\sqrt{4-\delta}$,
so that $\tanh 2 \phi_{k_{m}} \rightarrow 1$.
There are more choices for the order parameter at $h=h_c$ compared to either $h_{c1}$ or $h_{c2}$  
as both $\psi_{\pm}$ and ${\bar \psi}_{\pm}$ 
condense at $h_c$. The full expression for the ground state energy at $\delta =4$, $h=h_c$   
to fourth order in $\psi$ and ${\bar \psi}$, and to leading order in $1/S$ is
\begin{eqnarray}
\frac{2E}{JNS^2} &=& \left(|\psi_{+}|^2 + |\psi_-|^2 - |{\overline{\psi}}_+|^2 - |{\overline{\psi}}_-|^2\right)^2 \nonumber \\
&& + 2 |\psi_{+}|^2  |{\overline{\psi}}_+|^2 + 2 |\psi_-|^2 |{\overline{\psi}}_-|^2
\label{n_3}
\end{eqnarray}
subject to constraint $\psi_- \psi^*_{+} - {\overline{\psi}}_- {\overline{\psi}}^*_+ = 
i(\psi_-  {\overline{\psi}}_{-} + \psi^*_{+} {\overline{\psi}}^*_+)$ which eliminates 
infinitely large terms from the energy.  
Choosing just one of the four order parameters non-zero, we obtain the 
same distorted umbrella states as before, with $E \propto |\psi|^4$ and a finite chirality. 
However, we see from (\ref{n_3}) that
there is a better choice---taking $|\psi_+| = |{\overline{\psi}}_-| \neq 0$, 
 $\psi_- = {\overline{\psi}}_+ =0$ or vise versa, we find that $E =0$.
Thus, unlike the situation at any other point on the critical lines $h_{c1,2}(\delta)$,
 the magnitude of the condensate 
$|\psi_{+}| = |{\bar \psi}_{-}|$ at
the end-point of the UUD phase is unconstrained implying that $|\psi_{+}| =
 |{\bar \psi}_{-}|$ jumps to a finite value 
right at $\delta =4$.  
The chirality $K_{ABC}$ of  such a state depends on the relative phase $\theta$ of the two order parameters 
as  $ K_{ABC} \propto \tanh {2 \phi_{k_{m}}} + \sin \theta$ and
 vanishes when 
$\theta = -\arcsin(\tanh {2 \phi_{k_{m}}}) \rightarrow -\pi/2$.  
We verified that this particular $\theta$ is the only choice at which 
the transverse magnetization given by $\langle S^{x,y}_{A}\rangle =0$, 
$\langle S^+_{B}\rangle = - \langle S^+_{C}\rangle \sim |\psi_+| 
\sqrt{4-\delta}  ~e^{-ik_{\rm m} x}$  does not diverge 
together with the coherence factors 
but rather remains zero at $\delta = 4$,
 even though $|\psi_{+}| \neq 0$ there.  
 It is tempting to speculate that such a zero-chirality state persists  
beyond $\delta=4$ along a line in the $h-\delta$ plane
and continues to separate the two distorted umbrella phases
 (it cannot exist in a finite $h$-range since, unlike the UUD phase, it does not have
 two gapped low-energy modes).
The spin structure along this line is either collinear, as at $\delta = 4$,
or coplanar, with $\langle S^{x,y}_{A}\rangle =0$, $\langle S^+_{B}\rangle = - \langle S^+_{C}\rangle$;
the difference can not be resolved within our formalism. 

\emph{Energy considerations and the phase diagram.}~
 So far we have analyzed the UUD phase's local stability without
 addressing whether it globally minimizes the energy.
There are three regimes where one can easily compare the umbrella and planar energies.
First is the high field regime $h \approx h_{\text{sat}}$.
There, the umbrella state, which at arbitrary $h$ is described by
\begin{equation}
   {\bf S}_{\bf r} = S\{\cos\theta[\cos({\bf Q \cdot r}) {\bf \hat 
{x}} + \sin({\bf Q \cdot r}) {\bf \hat{y}}] + \sin\theta {\bf \hat{z}} \}
\label{nn_1}
\end{equation}
with  ${\bf Q} = 2 \cos^{-1}(-J'/2J)$ and $\sin\theta = h/h_{\rm sat}$,
wins for all $J' \neq J$ simply because quantum effects vanish at $h_{\text{sat}}$.  We have verified this explicitly by computing the analog of Eq.\ (\ref{n_1}) at the saturation field to show that indeed interactions drive the system into the umbrella state for arbitrary $J' \neq J$.  
As a result, the critical line which begins at $\delta =3$, $h = h_{c2}$ 
should end up at  $\delta =0$, $h= h_{sat}$.

The second regime occurs at small $h\to 0$.
Here the lowest-energy planar configuration is
{\it incommensurate}, with the same ${\bf Q}$ as the umbrella state and 
\begin{equation}
  {\bf S}_{\bf r} = S[\cos({\bf Q\cdot r} + \varphi_{\bf r}){\bf  
\hat{z}} + \sin({\bf Q\cdot r} + \varphi_{\bf r}){\bf \hat{x}}] ,
\label{nn_2}
   \end{equation}
where $\varphi_{\bf r} = - (2 h/u)\sin({\bf Q\cdot r}) + O(h^2)$
and  $u = h_{\rm sat}[1 + (J-J')^2/J^2]$.  
At small $h$, the energy difference between the 
incommensurate umbrella and planar states of Eqs.\ (\ref{nn_1}) and (\ref{nn_2}) is 
$\Delta E_{h \to 0} \equiv (E_{\rm umb}-E_{\rm pl})/NS^2 = -(1/2) h^2  \Delta \chi$, where 
$ \Delta \chi = \chi_{\rm umb}-\chi_{\rm pl}$ is the difference of susceptibilities.  
In the classical limit we find
$\chi_{\rm umb} = 1/h_{\rm sat}$, $\chi_{\rm pl} = 1/u$, so that $\Delta \chi = (J-J')^2/(9J^3)$ and
  the umbrella state has lower energy. 
The competition comes from quantum fluctuations: $1/S$ corrections to 
 $\chi_{\rm umb}$ and $\chi_{\rm pl}$ are different already for $J = J'$, and such  
 that $\Delta \chi_{qu} \approx -0.16/(18JS)$ (Ref.~\cite{chubukov91}). 
 Adding the two contributions, we find that $\Delta E_{h \to 0} = [0.008 h^2/(2J S)] (1.1-\delta)$, \emph{i.e.}, 
 the incommensurate planar state has lower energy for $\delta < 1.1$.
This implies that the commensurate planar state that 
we found immediately below $h_{c1}$ should undergo either a second- or first-order transition into an incommensurate 
planar state at some $h < h_{c1}$. 
We therefore expect the line separating planar and distorted umbrella states at 
low fields to depart at $\delta =1$, $h = h_{c1}$ and 
end up at $\delta =1.1$, $h =0$.

Finally, at $h_{\rm sat}/3$ the energy difference between the umbrella and UUD phase is
 $\Delta E_{1/3} = [0.067 J/(2S)] (2.0-\delta)$,
 where the first and second terms, respectively, are the classical and quantum contributions;
 see Ref.~\cite{chubukov91}.
Consequently, the UUD phase
and the neighboring distorted umbrella phases remain global minima only up to $\delta =2.0$ and become metastable at larger $\delta$. This suggests that
for $\delta >2$ the UUD state can be observed only via a transient
 magnetization plateau, similar to the situation in 
 a kagom\'e system \cite{narumi04}.
Equilibrium measurements should reveal only umbrella-like states in that region of $\delta$.

The resulting phase diagram near 1/3 magnetization is shown in Fig.\ \ref{PhaseDiagram}.  It contains an 
 UUD phase; two commensurate planar states from Fig.\ \ref{GroundStates}(c); and two 
non-coplanar incommensurate distorted umbrella phases.  The shaded region corresponds to the regime where the classical umbrella minimizes the energy globally.  Additionally, incommensurate planar states are expected at small fields when $\delta \leq 1$, and near the saturation field for small $\delta$.
 We also expect new  phases at 
 small  $J'/J$ (large $\delta$), where one-dimensional physics takes over the semi-classical analysis.  
 
This phase diagram is in agreement with data for \CsCuBr, where $J'/J = 0.7$ implies that $\delta = 0.6$ if we extrapolate to $S = 1/2$.  For this $\delta$, the UUD state is present, and the nearby phases are 
planar, in agreement with NMR
\cite{takigawa04,takigawa07} and neutron \cite{ono05} experiments.
These experiments also observe that both transitions out of the UUD state
are first order. Our calculations predict continuous transitions as a 
consequence of the U(1) spin symmetry exhibited by the Hamiltonian (\ref{spinH}).
However, when this U(1) symmetry is broken explicitly by spin-orbit coupling, cubic terms in the free energy 
are permissible, which generically render the transition first order. 
In particular, DM coupling of the form present in \CsCuBr~breaks this symmetry when the field is directed 
along the triangular layers.  In addition, a 
direct first order transition from UUD phase into 
 the incommensurate planar phase 
is also a possibility, which should be investigated by numerical calculations
similar to those in \cite{yoshikawa04}.
For  \CsCuCl, the anisotropy is much higher ($\delta \approx 2.9$), and the system very 
likely lies outside of the applicability region of our analysis, and should be approached from a 1D
perspective \cite{starykh07}.
Still, even within our framework, $\delta > 2$ implies no UUD phase, and
no plateau is seen in \CsCuCl. 

An intriguing question concerns the possible appearance of a  2/3-magnetization plateau at 
 $\delta <1$ and higher fields, as observed in \CsCuBr \cite{ono04}, which would correspond to (at least)
a ``5-up, 1-down'' configuration.  
While such states are never ground 
states to order $1/S$, we verified that their energy
 is reduced when $J' \neq J$.  
We speculate that, due to a large degeneracy of 5-up, 1-down configurations, 
a 2/3-plateau may be \emph{entropically} stabilized  at
finite temperature. Regarding this issue, the role of spin-phonon couplings should be seriously
investigated \cite{wang08}.

 \emph{Conclusions.}~
Using a modified large-$S$ approach, we studied the quantum phase diagram of an anisotropic triangular antiferromagnet, with particular emphasis on the classically unstable UUD state and proximate phases.  Fig.\ \ref{PhaseDiagram} summarizes our findings.
The UUD phase with $1/3$ magnetization plateau survives 
a substantial range of anisotropy, and at its boundaries transforms either into 
commensurate planar phases, or into umbrella-like incommensurate chiral
phases, depending on whether the spin-wave instabilities of the UUD phase are 
at zero or finite momenta.  Our results explain a number of experimental findings for \CsCuBr.  

\acknowledgments{We acknowledge illuminating discussions with O. Motrunich, 
M. Takigawa, and Y. Takano, who we also thank for sharing experimental data.  This work  
was supported by the NSF DMR-0210790 and
the Lee A.\ DuBridge Foundation (JA), and by 
 NSF-DMR 0604406 (A.V. Ch).}

\bibliography{UUD}

\begin{references}
\bibitem{veillette05} M. Y. Veillette, J. T. Chalker, and R. Coldea, \prb {\bf 71}, 214426 (2005).
\bibitem{tokiwa06} Y. Tokiwa  {\sl et al.}, \prb {\bf 73}, 134414 (2006).
\bibitem{veillette06} M.Y. Veillette and J.T. Chalker, \prb {\bf 74}, 052402 (2006).
\bibitem{ono03} T. Ono   {\sl et al.}, \prb {\bf 67}, 104431 (2003).
\bibitem{ono04}  T. Ono   {\sl et al.}, J. Phys.: Condens. Matter {\bf 16}, S773 (2004).
\bibitem{ono05}  T. Ono   {\sl et al.}, Prog. Theor. Phys. Suppl. {\bf 159}, 217 (2005).
\bibitem{takano07} H. Tsuji {\sl et al.}, \prb {\bf 76}, 060406(R) (2007).
\bibitem{takigawa04} Y. Fujii {\sl et al.}, Physica B {\bf 346-347}, 45 (2004).
\bibitem{takigawa07} Y. Fujii {\sl et al.}, J. Phys.: Condens. Matter {\bf 19}, 145237 (2007).
\bibitem{takano08} N. Fortune, S. Hannahs, Y. Yoshida, Y. Takano, T. Ono, and H. Tanaka, unpublished.
\bibitem{chubukov91} A. V. Chubukov and D. I. Golosov, J. Phys.: Condens. Matter {\bf 3}, 69 (1991). 
The lifting of the ``accidental'' degeneracy by classical, thermal fluctuations has 
been analyzed by  H. Kawamura and S. Miyashita, J. Phys. Soc. Jpn. {\bf 54}, 4530 (1985).
\bibitem{narumi04} Y. Narumi {\sl et al.}, Europhys. Lett. {\bf 65}, 705 (2004).
\bibitem{shender82} E.F. Shender, Sov. Phys. JETP {\bf 56}, 178 (1982).
\bibitem{henley88} C. L. Henley, \prl {\bf 62}, 2056 (1989).
\bibitem{honecker04} A. Honecker, J. Schulenburg, and J. Richter, J.Phys.: Condens. Matter {\bf 16}, S749 (2004).
\bibitem{nikuni95} T. Nikuni and S. Shiba, J. Phys. Soc. Jpn. {\bf 64}, 3471 (1995).
\bibitem{yoshikawa04} S. Yoshikawa {\sl et al.}, J. Phys. Soc. Jpn. {\bf 73}, 1798 (2004).
\bibitem{miyahara06} S. Miyahara, K. Ogino, and N. Furukawa, Physica B {\bf 378-380}, 587 (2006).
\bibitem{starykh07} O. A. Starykh and L. Balents, \prl {\bf 98}, 077205 (2007).
\bibitem{wang08} F. Wang and A. Vishwanath, \prl {\bf 100}, 077201 (2008).

\end{references}

\end{document}